\documentclass[sigconf,authorversion]{acmart}

\usepackage{booktabs}
\usepackage{amsmath}
\usepackage[normalem]{ulem}  
\usepackage{bbold}  
\usepackage{subfigure}  
\usepackage{placeins}  
\usepackage{multirow} 
\usepackage{moreverb} 
\usepackage{tabularx} 
\usepackage{upgreek}  

\newcommand{\upup}{$^{\blacktriangle}$}

\newcommand{\vect}[1]{\boldsymbol{#1}}

\newcommand{\sectionshrink}{\vspace*{-.3\baselineskip}}

\setcopyright{acmcopyright}

\copyrightyear{2019}
\acmYear{2019}
\acmConference[ICTIR '19]{The 2019 ACM SIGIR International Conference on the Theory of Information Retrieval}{October 2--5, 2019}{Santa Clara, CA, USA}
\acmPrice{15.00}
\acmDOI{10.1145/3341981.3344244}
\acmISBN{978-1-4503-6881-0/19/10}

\begin{document}

\title{Unsupervised Context Retrieval for Long-tail Entities} 


\author{Dar\'{i}o Garigliotti$^{*}$}
\thanks{$^{*}$ Work done while Dar\'{i}o Garigliotti was visiting Signal.}
\affiliation{University of Stavanger, Norway}
\email{dario.garigliotti@uis.no}

\author{Dyaa Albakour}
\affiliation{Signal, 145 City Road, London EC1V 1AZ, UK}
\email{dyaa.albakour@signal-ai.com}

\author{Miguel Martinez}
\affiliation{Signal, 145 City Road, London EC1V 1AZ, UK}
\email{miguel.martinez@signal-ai.com}

\author{Krisztian Balog}
\affiliation{University of Stavanger, Norway}
\email{krisztian.balog@uis.no}


\maketitle

\begin{abstract}
Monitoring entities in media streams often relies on rich entity representations, like structured information available in a knowledge base (KB).
For long-tail entities, such monitoring is highly challenging, due to their limited, if not entirely missing, representation in the reference KB.
In this paper, we address the problem of retrieving textual contexts for monitoring long-tail entities.
We propose an unsupervised method to overcome the limited representation of long-tail entities by leveraging established entities and their contexts as support information.
Evaluation on a purpose-built test collection shows the suitability of our approach and its robustness for out-of-KB entities.
\end{abstract}



\section{Introduction}
\label{sec:intro}

Entities, like people and organizations, are at the core of multiple techniques and applications devised to automatically process text documents~\citep{Farber:2016:EED, Graus:2018:BCM, Spitz:2018:EEN}.
One of these applications is reputation analysis, for example, in monitoring through news streams the stock performance of a company or a scandal around a celebrity.
This task, termed \emph{online reputation management}~\citep{Amigo:2012:ORL}, is concerned with identifying and mining textual contexts from media streams, e.g.\ sentences in news articles, in which a given entity occurs.
In order to identify such contexts, it is necessary to detect mentions of entities and disambiguate them, making this problem closely related to that of \emph{entity linking} (EL)~\citep{Rizzo:2017:LLN,Erp:2016:EEL,Esquivel:2017:OLT,Ilievski:2018:SSL}.
Entity linking is a prime example of entity-centered problems that are naturally addressed with methods that rely on knowledge bases (KBs)---large-scale repositories containing properties and facts about entities in a structured format.
The suitability of such semantics-aware methods, nevertheless, is limited by the amount of information available for a given entity.
EL benchmarks, for example, usually report very good performance, however, the evaluation datasets are focused on the head of the entities distribution (i.e., popular entities), and systems are trained to perform well on those entities~\citep{Ilievski:2018:SSL}.
Conversely, entities in the \emph{long tail} are out of focus for most entity-oriented approaches~\citep{Erp:2016:EEL}.  

%
\begin{figure}[t]
	\vspace{0.25in}
	\centering
	\hspace{-0.1in}	\includegraphics[width=0.485\textwidth]{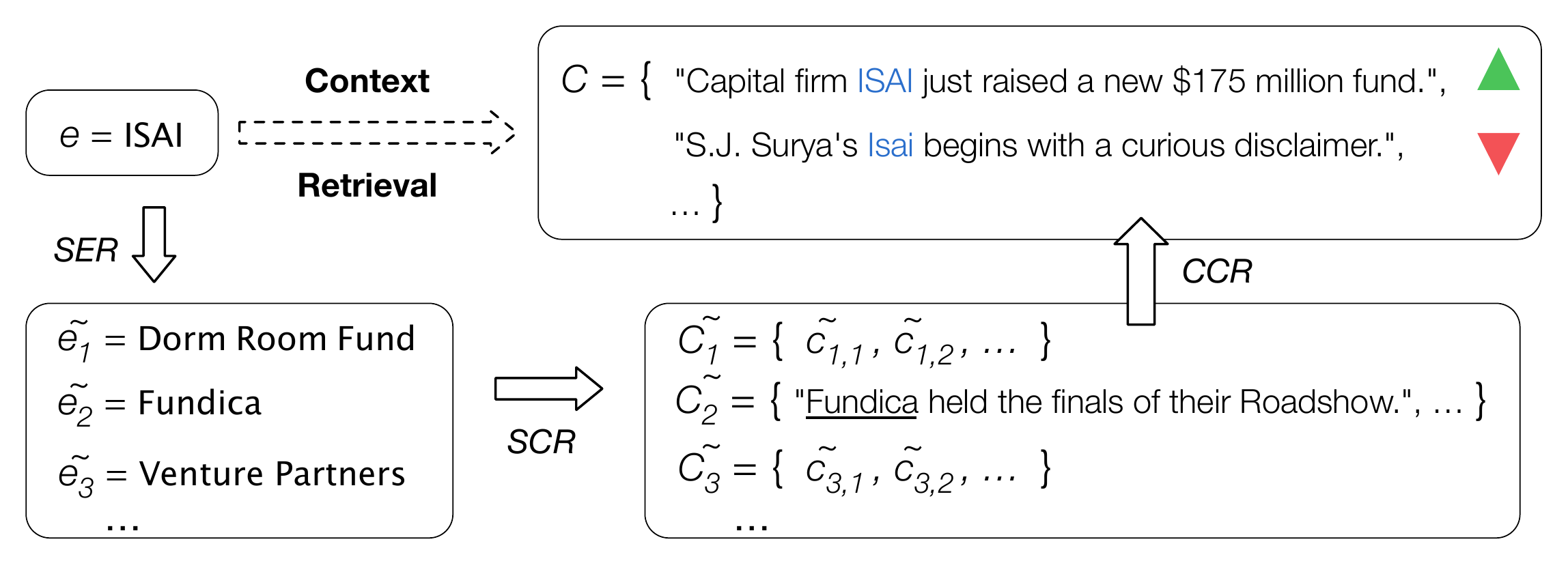}
    \caption{ 
        Overview of our proposed framework.  
    }
	\label{fig:overview}
	\vspace{-0.32in}
\end{figure}

In this paper, we address the problem of identifying contexts for monitoring long-tail entities in news articles.
Specifically, we propose a generative probabilistic framework to rank contexts where a surface form of a long-tail entity occurs, by leveraging support information.
As shown experimentally, our approach is in particular able to cope with out-of-KB entities in a robust manner.

To illustrate the main ideas behind our framework, consider Fig.~\ref{fig:overview}, where we aim to monitor the long-tail input entity $e = \texttt{ISAI}$, a French enterpreneurs' fund.
Then, we wish to identify contexts (here: sentences) that mention one of the known surface forms of $e$ (e.g., ``ISAI'' or ``Isai'').
It is often the case that two or more entities share the same surface form.  
For example, in the context ``Capital firm \textbf{ISAI} just raised a new \$175 million fund,'' the surface form indeed refers to \texttt{ISAI}, meanwhile in ``S.J. Surya's \textbf{Isai} begins with a curious disclaimer,'' it refers to an Indian movie.
Deciding which entity a given mention refers to is known as entity disambiguation, a well-studied subtask in the context of entity linking.  Modern entity disambiguation approaches leverage rich semantic information associated with entities in a KB~\citep{Balog:2018:EOS}---information which is lacking, or missing altogether, for long-tail entities.
This renders supervised learning methods unsuitable and motivates us to develop an unsupervised solution.
Our main underlying idea is to utilize established entities (referred to as \emph{support entities}, with rich KB entries) similar to the input entity, and their contexts (\emph{support contexts}), to rank the contexts in which the input entity is mentioned.  
Therefore, we first rank support entities (SER step in Fig.~\ref{fig:overview}) according to how similar they are to $e$,  e.g., $\tilde{e} = \texttt{Fundica}$.  Next, we identify support contexts for each support entity via traditional entity linking (SCR step in Fig.~\ref{fig:overview}), e.g., $\tilde{c} =$ ``Fundica held the finals of their Roadshow.''
Finally, we rank contexts for our input entity by considering their similarity to the support contexts (CCR step in Fig.~\ref{fig:overview}).

For evaluation, we build a curated test collection of 165 long-tail entities (comprised of 92 in-KB entities with Wikipedia pages and 73 out-of-KB entities) and collect relevance assessments for contexts retrieved from a large news corpus.
We show experimentally that our method substantially outperforms both an entity linking and a retrieval baseline.  
Moreover, it performs just as well for entities without any representation in Wikipedia, a setting where the entity linking baseline fails inevitably.  
The contributions that we make in this work are: (i) an unsupervised context retrieval approach for long-tail entities, (ii) a purpose-built test collection, and (iii) a detailed component-level evaluation and an analysis of performance for in-KB vs. out-of-KB entities.  
\section{Problem Statement}
\label{sec:problem}

We first clarify some terminology and assumptions.

Each entity $e$ in a knowledge base (KB) has a unique identifier.
A \emph{surface form}, or \emph{alias}, of $e$ is a textual phrase used to refer to $e$.
For example, ``Obama'' or ``US President'' may be some of the known aliases to refer to the entity whose
identifier is $\texttt{Barack\_Obama}$.
An entity \emph{mention} is a surface form occurring in some context (span of text).
For simplicity, in this paper we take contexts to be sentences, i.e., the unit of retrieval are sentences.

In the rest of this work, $e$ is a long-tail entity, and $C$ is a set of contexts each with at least one alias of $e$ being mentioned.
We assume that we know (i) a brief textual description $e_{desc}$, (ii) the entity type (among Person, Location, or Organization), and (iii) all the entity surface forms.
These assumptions are intuitive in our scenario of entity monitoring in media, where all this information can be easily traced and provided by an external component.
We also assume that we have access to a large collection of entities $\mathcal{E}$ from an entity catalog or a KB, e.g.\ Wikipedia.  
Another assumption is that all entity mentions in the contexts are detected.

Given $e$ and $C$, we aim to rank each context $c \in C$ according to the likelihood that the (potentially ambiguous) entity mention in $c$ refers to $e$.
The context retrieval problem then consists in assigning a score $s(c, e)$ to each pair $(c, e)$ indicating our confidence that $e$ is the entity mentioned in $c$.
\section{Generative Modeling Framework}
\label{sec:approach}

We introduce a generative probabilistic model for scoring a context $c \in C$ according to $P(c|e)$, i.e., the probability that the surface form mentioned in $c$ refers to the long-tail entity $e$.
Formally:

\small
\begin{align}
	P(c|e) & = \sum_{\tilde{e}} P(c|e, \tilde{e}) P(\tilde{e}|e) \nonumber \\
	& = \sum_{\tilde{e}} \big(\sum_{\tilde{c}} P(c|e, \tilde{e}, \tilde{c}) P(\tilde{c}|e, \tilde{e})\big) P(\tilde{e}|e)  \nonumber \\
	& = \sum_{\tilde{e}} \big(\sum_{\tilde{c}} P(c|e, \tilde{c}) P(\tilde{c}|\tilde{e})\big) P(\tilde{e}|e) \label{eq:exp_3} \\
\nonumber
\end{align}
\normalsize

This model has three components (illustrated in Fig.~\ref{fig:overview}):

(i) $P(\tilde{e}|e)$ (\emph{Support Entity Ranking}, or SER) expresses the importance of a \emph{support} entity $\tilde{e}$ for the given long-tail entity $e$;

(ii) $P(\tilde{c}|\tilde{e})$ (\emph{Support Context Ranking}, or SCR) represents the importance of a \emph{support} context $\tilde{c}$ for a support entity $\tilde{e}$; and

(iii) $P(c|e, \tilde{c})$ (\emph{Context-to-Context Ranking}, or CCR) is the importance of a support context $\tilde{c}$ for a context $c$, given that an alias of the long-tail entity $e$ is mentioned in $c$.

The last expression in Eq.~\eqref{eq:exp_3} is obtained after assuming independence of $c$ w.r.t. $\tilde{e}$ given $\tilde{c}$, and of $\tilde{c}$ w.r.t. $e$ given $\tilde{e}$.

\subsection{Support Entity Ranking}
\label{sec:approach:ee_ranking}

Since $e$ has a limited representation in KBs and in contexts where its known aliases might be mentioned, the textual description $e_{desc}$ we assume known (cf. Sect.~\ref{sec:problem}) is key to find relevant entities.
It is appropriate then to consider a text matching method that ranks support entities for $e$.
This method can be, for example, a retrieval model that uses $e_{desc}$ to query the entity collection $\mathcal{E}$.
This collection should contain mostly entities with good data representation, e.g, by being taken from an actual KB.

Given an entity ranking function $ser$, we denote the importance score of $\tilde{e}$ for $e$ by $ser(e, \tilde{e})$.
We can then estimate $P(\tilde{e}|e)$ by $ser(e, \tilde{e}) / \sum_{e' \in \tilde{E}}ser(e, e')$.

\subsection{Support Context Ranking}
\label{sec:approach:ec_ranking}

Once $N$ support entities $\tilde{E} = (\tilde{e}_1, ..., \tilde{e}_N)$ are ranked for $e$, we aim to exploit the contexts where it is known that entities from $\tilde{E}$ are mentioned.
Specifically, for each $\tilde{e}_i, i \in {1, ..., N}$, let $\mathcal{C}_{\tilde{e}_i}$ be a collection of contexts, such that for each $\tilde{c_x} \in \mathcal{C}_{\tilde{e}_i}$, $\tilde{c_x}$ contains a mention \emph{linked} to $\tilde{e}_i$.
We assume that each of these collections is previously built, by applying an entity linking system over a universe of context candidates.
Then, we take $M_i$ support contexts $\tilde{C_i} = \{\tilde{c}_1, ..., \tilde{c}_{M_i}\}$ among all the contexts in  $\mathcal{C}_{\tilde{e}_i}$.
For each $\tilde{C_i}$, its size $M_i$ would be informative enough, since the ``support'' entity collection $\mathcal{E}$ is mostly head-oriented, and there the EL system should link each $\tilde{e}_i$ from a sufficiently large number of contexts.
We refer to the contexts of any $\tilde{C}$ as \emph{support contexts}, and say that the previously applied EL method links an entity mention in $\tilde{c}$ to $\tilde{e}$ with confidence $conf(\tilde{c}, \tilde{e})$.
The set of all support contexts for the long-tail entity $e$ is defined as $\tilde{C} = \bigcup_i \tilde{C}_i$.
We can then estimate $P(\tilde{c}|\tilde{e})$ with $conf(\tilde{c}, \tilde{e}) / \sum_{c' \in \tilde{C}}conf(c', \tilde{e})$.

\subsection{Context-to-Context Ranking}
\label{sec:approach:cc_ranking}

Given $e$ and a context $c \in C$, $ccr(c, \tilde{c})$ denotes the relevance score of $c$ for each support context $\tilde{c} \in \tilde{C}$.
In order to estimate $P(c|e, \tilde{c})$, we define the following estimator: $ccr(c, \tilde{c}) / \sum_{c' \in C}ccr(c', \tilde{c})$.
\section{Experimental Setup}
\label{sec:expers}

This section describes the estimators, dataset, evaluation metrics and baselines that we use in our experiments.  

\emph{Component estimators.}  For estimating the SER component (cf. Sect.~\ref{sec:approach:ee_ranking}), we take articles from Wikipedia (2018-10-01 dump) to be our entity collection $\mathcal{E}$.  
We then approach entity ranking via a standard unstructured retrieval model, specifically, BM25~\citep{Robertson:2009:PRF} with parameters set as $k1 = 1.2$ and $b = 0.8$, as recommended in~\citep{Hasibi:2017:DVT}.  
The entity description $e_{desc}$ is used as a query over the \texttt{opening\_text} field to rank at most 200 support entities.  
We refer to this SER definition as \emph{basic}.  
We also consider two variants to this method: (i) \emph{pop}, in which the basic retrieval score is multiplied by the popularity of the support entity estimated by the number of incoming links of its Wikipedia article; and (ii) \emph{types}, where we filter out, from the ranking, entities which do not have common types (assigned from the DBpedia Ontology) with the type of $e$; t\emph{types} corresponds to the strict filtering method in~\citep{Garigliotti:2017:OTE}.  
Type information has been shown to significantly improve retrieval performance, e.g., in the scenario of ad hoc entity retrieval~\citep{Garigliotti:2019:IET}.  
A parameter $N \in \{50, 100\}$ controls the cut-off of the support entity ranking.  

Secondly, we describe the SCR estimation (cf. Sect.~\ref{sec:approach:ec_ranking}).  
We work with a sample of 5 million news articles, published between October 2017 and September 2018, provided by commercial news streams aggregators.
We perform entity linking on the content of each article, using Spotlight~\citep{Daiber:2013:IEA} version 1.0.0 (with the latest model released for English language: 2016-10)\footnote{\url{https://sourceforge.net/projects/dbpedia-spotlight/files/2016-10/en/model/en.tar.gz/download}}.    
For each entity mention in an article, we retain an entity linking candidate only if its final confidence score is above a threshold $\theta$ of 0.9, to ensure high-quality support information.  
Then, from each article, we pick a single context (sentence) with an entity mention that has at least one retained candidate, regardless which entity it is.  
Given a support entity $\tilde{e}$, $\mathcal{C}_{\tilde{e}}$ is the set of all contexts where $\tilde{e}$ was linked, and so we take $\tilde{C}$ for $\tilde{e}$ as any non-empty subset of $\mathcal{C}_{\tilde{e}}$.  
A parameter $M \in \{50, 100\}$ controls the size of $\tilde{C}$.  

Finally, regarding the estimation of the CCR component (cf. Sect.~\ref{sec:approach:cc_ranking}), we consider two approaches to model $ccr$.  
Firstly, we define $ccr(c, \tilde{c})$ to be the retrieval score (BM25, $k1 = 1.2, b = 0.8$) of $c$ when using $\tilde{c}$ as a query over an index of contexts.  
We refer to this method as $retrieval$.  
Alternatively, we model $ccr(c, \tilde{c})$ as a semantic similarity between the long-tail entity context $c$ and a support context $\tilde{c}$.  
Specifically, we define a $semantic$ method to be the cosine similarity between the vectors $\vect{v}_c$ and $\vect{v}_{\tilde{c}}$, where $\vect{v}_s$ is the average, for all the terms in a context $s$, of term vectors in the pre-trained \emph{word2vec} word embeddings~\citep{Mikolov:2013:DRW}.

\emph{Test collection.}  In order to ensure quality in the earliest step, we manually identify a set $E$ of 165 long-tail entities.  
These correspond to either entities that have recently emerged, or entities with certain lifespan yet presenting difficulties to be linked with reasonable accuracy.  
162 out of the 165 entities (i.e., 98.18\%) are of type Organization, and 3 are people.  
After requesting for responses to the Wikipedia RESTful API\footnote{\url{https://en.wikipedia.org/w/api.php}}, we find that 73 out the 165 entities (i.e., 44.24\%) ---all organizations--- do not have a corresponding Wikipedia page by October 1st, 2018.  
For this subset of entities, the average number of known surface forms is 1.07 per entity; for the remainder 92 in Wikipedia, the average is 1.11 per entity.  

For each long-tail entity $e \in E$, we search for contexts where at least one of its surface forms occurs.  
Specifically, for a given $e$, we rank a set $C$ of at most 5,000 contexts from a proprietary collection using each of the possible \emph{configurations} (i.e., combinations of estimators and parameter settings for SER, SCR and CCR components; cf. Sect.~\ref{sec:approach}).  
We use the well-established top-$k$ pooling technique~\citep{Voorhees:2014:ESS} for our relevance assessment.  
We take the top 20 contexts from each of these rankings, and build a pool.  
Each of these 4,536 pooled contexts is assessed by an expert media analyst regarding whether it is relevant or not for its long-tail entity.  

\emph{Evaluation Metrics.}  We use two evaluation metrics: Mean Average Precision (MAP) and Mean Reciprocal Ranking (MRR), both very sensitive to highly ranked relevant contexts.  

\emph{Baselines.} Our baseline, a strong sentence retrieval model~\citep{Blanco:2010:FSS}, ranks the contexts in $C$ for a long-tail entity $e$ by the standard BM25 retrieval score of querying with $e_{desc}$ an index of $C$.\footnote{We also experimented with a combined method that mixes entity frequency and rarity~\citep{Blanco:2010:FSS}, but it performed worse than our BM25 baseline.  
It appears that noise is introduced when counting mentions of ambiguous aliases in contexts.}  
(Note that this is different from the $retrieval$ CCR component described above.)    
We also apply the Spotlight entity linker, previously used for estimating SCR (cf. Sect.~\ref{sec:approach:ec_ranking}), over each context in the universe for all long-tail entities, $\bigcup_e C_e$.  
We experiment with two values for the Spotlight final confidence threshold $\theta$: 0.6, as the best performing Spotlight setting~\citep{Mendes:2011:DSS}, and 0.9 as before.

\section{Experimental Results}
\label{sec:results}

Our experiments address three research questions: (RQ1) What is the best way to estimate each component?, (RQ2) How does our approach perform for context retrieval?, and (RQ3) How does it perform for entities with and without representation in a KB?

\subsection{Framework Components}
\label{sec:results:estimators}

\begin{table}[t]
  \small
  \centering
  \caption{Context retrieval performance, measured in terms of MAP and MRR metrics.
  Best values per block are typeset in bold.
  Statistical significance is tested using a two-tailed paired t-test at $p<0.05$ ($^\dag$) and $p<0.001$ ($^\ddag$).  
  For each line in the $pop$ and $types$ blocks, significance is tested against the corresponding row of the $basic$ block.  
  }
  \begin{tabular}{ l r r l l l l }
    \toprule
    & & & \multicolumn{4}{c}{$ccr$} \\
    \cline{4-7}
    $ser$
     & $N$
     & $M$
     & \multicolumn{2}{c}{semantic} & \multicolumn{2}{c}{retrieval} \\
    & & & MAP & MRR & MAP & MRR \\
    \toprule
	basic
	& 50 & 50
	& \textbf{0.5195} & 0.6133
	& 0.2758 & 0.4440 \\
	& 50 & 100
	& 0.5076 & 0.6138
	& 0.2796 & 0.4453 \\
	& 100 & 50
	& 0.5133 & \textbf{0.6357}
	& 0.2927 & 0.4449 \\
	& 100 & 100
	& 0.5081 & 0.6341
	& 0.2954 & 0.4489 \\
    \midrule
	pop
	& 50 & 50
	& 0.4441$^\ddag$ & 0.5738$^\dag$
	& 0.2636 & 0.4030 \\
	& 50 & 100
	& 0.4499$^\ddag$ & 0.5738
	& 0.2695 & 0.3826$^\dag$ \\
	& 100 & 50
	& 0.4564$^\ddag$ & 0.6061$^\dag$
	& 0.2683$^\dag$ & 0.3988$^\dag$ \\
	& 100 & 100
	& \textbf{0.4633}$^\ddag$ & \textbf{0.6200}
	& 0.2706$^\dag$ & 0.4067$^\dag$ \\
    \midrule
	types        
	& 50 & 50
	& 0.4512$^\ddag$ & 0.5818
	& 0.2537 & 0.3942 \\
	& 50 & 100
	& 0.4475$^\ddag$ & 0.5818
	& 0.2558 & 0.3964 \\
	& 100 & 50
	& \textbf{0.4544}$^\ddag$ & 0.5866$^\dag$
	& 0.2504$^\dag$ & 0.3828$^\dag$ \\
	& 100 & 100
	& 0.4477$^\ddag$ & \textbf{0.5881}$^\dag$
	& 0.2572$^\dag$ & 0.3913$^\dag$ \\
    \bottomrule
  \end{tabular}
  \label{table:results-all_scorers}
  \vspace*{-0.8\baselineskip}
\end{table}

Our probabilistic model (cf. Sect.~\ref{sec:approach}) consists of three components: (i) Support Entity Ranking (SER), estimated by defining the function $ser$ using either basic, pop or types, and with cut-off parameter $N$; (ii) Support Context Ranking (SCR), estimated via entity linking confidence, with parameter $M$ for the size of the support contexts set; and (iii) Context-to-Context Ranking (CCR), modeled with retrieval or semantic estimator to define the function $ccr$.  

Table~\ref{table:results-all_scorers} presents the performance of every possible configuration for context retrieval.  
In order to answer (RQ1: what are the best component estimators?), we first notice that SER defined as \emph{basic} outperforms its \emph{pop} and \emph{types} variants. 
It does that for each of both CCR settings, and for any pair of values for the ($N, M$) parameters.   
The differences in favor of SER are highly significant in terms of MAP, and less or not significant according to MRR.  
Filtering by common type should help retain relevant support entities, but the results suggest that, even with slight differences in terms of MRR, the performance of the whole ranking in terms of MAP is degraded.  
Secondly, there is not a clear pattern in the performance when comparing settings for the $M$ parameter.  
Regarding the last component, the semantic definition of CCR leads to the best estimator.  
This shows that the vocabulary mismatch affects CCR, a gap against which the semantic method is more robust.

\subsection{Unsupervised Context Retrieval}
\label{sec:results:vs_baseline}

Moving to address RQ2, we compare, against the baseline, the best performing configuration of our approach, found in Sect.~\ref{sec:results:estimators} (i.e., using basic SER, $N = 50$, $M = 50$, and semantic CCR).  
The results in Table~\ref{table:results-best_vs_baselines} corresponding to the entire set $E$ of long-tail entities show that our method outperforms the baseline with high significance.  
This validates the suitability of our framework, that leverages support information to rank contexts, for long-tail entities.   
The very low performance of Spotlight 2016-10 confirms that most of the long-tail entities that happen to exist by October 2018 were indeed recently added between those two dates.

\subsection{Out-of-KB Entities}
\label{sec:results:out_of_WP}

To answer RQ3, we add the two last blocks in Table~\ref{table:results-best_vs_baselines} that present the retrieval performances in the two disjoint subsets of $E$, namely with and without a corresponding Wikipedia article.  
We observe that our approach not only outperforms the baseline highly significantly in each of both subsets, but is also robust for the subset of entities that are not present in Wikipedia.  
It is in this subset of entities, clearly long-tail, that our approach performs best.  

\begin{table}[t]
  \footnotesize
  \centering
  \caption{Context retrieval performance, comparing each of the baselines and the best configuration for our approach (according to MAP; cf. Table~\ref{table:results-all_scorers}).  
  For a given metric, the symbol \upup denotes statistical significance against the baseline, tested using a two-tailed paired t-test at $p<0.001$.         
  }
  \begin{tabular}{ l | l@{~} l@{~} | l@{~} l@{~} | l@{~} l@{~} }
    \toprule
	Entities
	& \multicolumn{2}{c|}{All}
	& \multicolumn{2}{c|}{In Wikipedia}
	& \multicolumn{2}{c}{Not in Wikipedia}
	\\
    \midrule
    Method
    & MAP & MRR
    & MAP & MRR
    & MAP & MRR
    \\
    \toprule
    Spotlight, $\theta = 0.6$
    & 0.0406 & 0.0826
    & 0.0726 & 0.1475
    & 0.0004 & 0.0008
    \\
    Spotlight, $\theta = 0.9$
    & 0.0393 & 0.0811
    & 0.0701 & 0.1447
    & 0.0004 & 0.0008
    \\
    \midrule
    Baseline
    & 0.2248 & 0.3357
    & 0.1732 & 0.2787
    & 0.2899 & 0.4077
    \\
    \midrule
    Our method
    & \textbf{0.5195}\upup & \textbf{0.6133}\upup
    & \textbf{0.4900}\upup & \textbf{0.6032}\upup
    & \textbf{0.5565}\upup & \textbf{0.6261}\upup
    \\
    \bottomrule
  \end{tabular}
  \label{table:results-best_vs_baselines}
  \vspace*{-1.2\baselineskip}
\end{table}
%
%
\section{Related Work}
\label{sec:rel}

\citet{Blanco:2010:FSS} are the first to formalize and evaluate the task of finding sentences to support the relationship between an entity and an ad-hoc query.  
In our task, we relax the query-aware constraint by only ranking contexts for an entity.  
Another difference is that we require for a context to contain an entity mention.  
Closely developed with production-scale media monitoring, recent related work has addressed a number of applications like tracking entity popularity in media~\citep{Graus:2018:BCM}, and entity-centric approaches for topic modeling~\citep{Spitz:2018:EEN}.  
\citet{Farber:2016:EED} formalize a series of challenges on missing entities and/or surface forms in KBs for entity linking, focusing on the detection of emerging entities in news articles.  
Our work faces all these long-tail entity challenges, i.e., whether or not a surface form matches any known surface form for the entities in the KB, and whether or not the correct entity actually exists in the KB.  
\citet{Reinanda:2016:DFL} introduce EIDF, an entity-independent, supervised learning model to retain documents in news that are vital to enhance a textual entity profile.  
Our task concerns with much smaller semantic units (sentences) as contexts for entities.  
Then, in our setting, an unsupervised approach appears more suitable, since structural document features they use, like informativeness or timeliness, are no longer able to be captured.

Supervised learning models for EL highly depend on fine-tuned training data from KBs during the candidate selection stage~\citep{Rizzo:2017:LLN}.
Related work also finds that EL datasets are skewed towards popular and frequent entities of general-purpose KBs, and emphasizes the need for datasets that focus on the long tail~\citep{Erp:2016:EEL}.  
In this line, a very large proportion of entities lack of Wikipedia coverage, leading to a long tail of entities unlinkable by state-of-the-art systems~\citep{Esquivel:2017:OLT}.  
Very recently, a systematic study verifies that EL systems are heavily optimized on head entities, yet perform worst for infrequent, highly ambiguous entities~\citep{Ilievski:2018:SSL}.  

\section{Conclusions}
\label{sec:concl}

We have presented an unsupervised approach to retrieve contexts for long-tail entities. 
We built a test collection, consisting in a set of long-tail entities manually identified, and relevance assessments for contexts.  
The experimental results show that our approach outperforms a text matching baseline, and also is robust for entities without representation in Wikipedia.   

In a scenario of a newly emerged entity under media management, the entity will likely lack a KB entry during a long period~\citep{Graus:2018:BCM}.  
Our method can be used in such cases to retrieve contexts, powering entity mining in news articles.
We see at least two additional applications of our approach.
Firstly, the top-ranked contexts identified by our approach could be used as weakly labeled instances in training entity linking models by distant supervision.
Secondly, these contexts could be used for augmenting textual entity profiles, e.g., by leveraging information from filtered news articles~\citep{Reinanda:2016:DFL} and by incorporating media streams in the entity representation~\citep{Graus:2016:DCE}.  

\FloatBarrier  
\sectionshrink
\bibliographystyle{ACM-Reference-Format}
\bibliography{long_tail}

\end{document}